\documentclass[english,reprint,amsmath,amssymb]{revtex4-1}
\usepackage[latin9]{inputenc}
\usepackage{babel}
\usepackage{mathtools}
\usepackage{amsmath}
\usepackage{amssymb}
\usepackage{mathdots}
\usepackage{graphicx}
\usepackage{esint}
\usepackage[unicode=true,
 bookmarks=false,
 breaklinks=false,pdfborder={0 0 1},backref=section,colorlinks=false]
 {hyperref}

\makeatletter

\usepackage{amsthm}
\usepackage{amsfonts}
\usepackage{mathrsfs}
\usepackage{dcolumn}
\usepackage{bm}

\makeatother

\begin{document}
\title{Ultrametric theory of conformational dynamics of protein molecules
in a functional state and the description of experiments on the kinetics
of CO binding to myoglobin}
\author{A.Kh.~Bikulov, }
\thanks{Semenov Institute of Chemical Physics, Russian Academy of Sciences,
ul. Kosygina 4, 117734 Moscow, Russia. E-mail: bikulov1903@rambler.ru.
Tel. +79166936434 }
\author{\\
 A.P.~Zubarev }
\thanks{Samara University, Moskovskoe shosse 34, Samara, 443123 Russia; Samara
State University of Railway Transport, Pervyi Bezymyannyi pereulok
18, Samara, 443066, Russia. E-mail: apzubarev@mail.ru. Tel. +79171081131}
\begin{abstract}
The paper is devoted to a systematic account of the theory of conformational
dynamics of protein molecules. As an example of application of this
theory, we provide a complete analytical description of experiments
on the kinetics of CO binding to myoglobin, which were carried out
by the group of Frauenfelder more than 30 years ago and acquired the
status of base experiments for studying the properties of the fluctuation
dynamic mobility of protein molecules. As early as 2001, the authors
could demonstrate that, within the model of ultrametric random walk
with a reaction sink, the experimental curves of CO binding to myoglobin
can be reproduced in the high-temperature region. Later, in 2010,
the authors proposed a modified model and, based on its numerical
analysis, demonstrated that this model can reproduces the experimental
results over the whole temperature range covered in the experiment.
In the present study, based on the previously proposed model, we formulate
a rigorous mathematical theory of conformational dynamics of protein
molecules. We demonstrate that the proposed theory provides not only
a complete description of the experiment over the whole temperature
range of $\left(60\div300\right)$ K and in the observation time window
of $\left(10^{-7}\div10^{2}\right)$ s but also a unified picture
of the conformational mobility of a protein molecule, as well as allows
one to realize the fact that the mobility changes in a self-similar
way. This specific feature of protein molecules, which has remained
hidden to date, significantly expands the ideas of dynamic symmetry
that proteins apparently possess. In addition, we show that the model
provides a prediction for the behavior of the kinetic curves of the
experiment in the low-temperature range of $\left(60\div180\right)$
K at times not covered by the experiment (more than $10^{2}$ s).
\end{abstract}
\keywords{kinetics of CO binding to myoglobin, ultrametricity, $p$-adic model
of conformational dynamics, $p$-adic mathematical physics}
\maketitle

\section{Introduction}

According to the ``protein-machine'' concept \citep{B}, protein
does not speed up but, conversely, slows down elementary chemical
acts (the formation or breaking of a chemical bond, charge transfer,
and so on). The slow conformational dynamics of a protein molecule
along a distinguished degree of freedom controls the behavior of an
elementary act in the active center of a protein molecule on a large
time scale from nanoseconds to hundreds of milliseconds and plays
the key role in the fermentative function \citep{B,BG,TM}. This feature
imparts to protein the properties of a ``machine'' that can manipulate
individual charges, atoms and molecules, against the background of
thermal fluctuations (the slower protein works, the lower the probability
of error). This leads to the necessity of studying conformational
dynamics on large time scales. Taking into account this concept, Frauenfelder
et al. \citep{ABB,SAB} carried out experiments on photodissociation
and subsequent binding of a CO molecule to myoglobin. In these experiments,
the authors obtained nontrivial results, which, unfortunately, could
not be consistently explained within the existing theories. To explain
the conformational dynamics of protein, Frauenfelder put forward the
idea of the ultrametricity of the energy landscape -- the hypersurface
of the potential energy of a protein molecule. These ideas were suggested
by the studies of Parisi \citep{Parisi} and Ramal, Toulouse, and
Virasoro \citep{RTV}. It is interesting that recent studies on this
subject -- the 2010 microreview by Frauenfelder \citep{F2010} and
the 2015 paper by K. Nienhaus and G. Nienhaus \citep{NN} -- remained
at the previous level of theoretical understanding. As a result, an
opinion has been formed about the kinetics of CO binding to myoglobin
that the states of myoglobin at physiological and low temperatures
differ quite significantly and that it is impossible to construct
a unified description of the binding kinetics within a single physical
model in the temperature range from $60$ to $300$ K. As for the
ultrametricity of conformational states of protein, in fact, this
idea dropped out of discussion (see, for details, our paper \citep{ABZ_2014},
where we discuss in detail the model formulas proposed for these experiments
and explain their problematic aspects).

Possibly, it is the lack of theoretical understanding that is responsible
for the eventual fading of interest in experiments of this kind. The
emergence of good femtosecond experimental setups turned the interest
of researchers to the study of small regions of protein molecules
(as a rule, the active center of protein) on femtosecond scales. It
is this period when interest in the models of small fragments of protein
molecules and the computer simulation of these models arose. Such
an approach leads to certain success and, for a proper choice of potentials,
may give good agreement with experiments (see, for example, \citep{Daw1,Daw2,Daw3}).
Nevertheless, it is worth noting that it is configurational rearrangements
of sufficiently large fragments of the system, including tens and
hundreds of elements, that play an important role in the behavior
of complex systems such as protein. As pointed out in \citep{Sher},
in complex systems, a ``conflict'' between local interactions and
the constraints imposed on the system (frozen bonds) gives rise to
strong ruggedness of the energy landscapes. If $N$ is the number
of elements taking part in the configurational rearrangements, then,
according to the estimates of \citep{Stillinger1,Stillinger2}, the
number of local minima of the energy landscape is on the order of
$\thicksim N!\exp\left(\eta N\right)$, where $\eta$ is a quantity
on the order of unity. It is such cases, which are most interesting
from the physical viewpoint, for which the full description of energy
landscapes and, hence, the study of the dynamics of the system becomes
impossible due to the unfeasibility of computations. For example,
a full computer reconstruction of the fundamental act of protein in
which the motions of all fragments of the protein structure would
be equally represented at times from $10^{-9}$ to $10^{0}$ s is
impossible. Thus, to date, irrespective of the computational resources,
the computer simulation of the conformational dynamics of structures
comparable with proteins in complexity is limited. It can, at best,
give either a relatively detailed idea about the behavior of a structure
in a relatively small domain of the conformational space, or (in a
strongly coarsened description) a sketchy representation of its behavior
as a whole.

Fundamental difficulties of this kind force us to seek fundamentally
new approaches to the description of the conformational dynamics of
protein. One of such approaches, which we would like to highlight,
was proposed in the works of Stillinger and Weber \citep{Stillinger3}
and Becker and Karplus \citep{BK}. Although this approach was initially
developed for computer simulation, it is nevertheless close to the
ideas of Frauenfelder on ultrametricity in proteins. This approach
is based on the representation of multidimensional strongly rugged
landscapes by hierarchical graphs. The authors of \citep{Stillinger3,BK}
applied the procedure of hierarchical clustering of the local minima
of the energy landscape. The procedure of clustering of local minima
was applied to the energy barriers between these minima. As a result
of this procedure, the local minima are combined into sets hierarchically
nested into each other, which are called ``basins.'' Such hierarchical
clustering (or, in other words, the taxonomy problem) resulted in
a treelike hierarchical structure of quasi-equilibrium states of protein.
Here the dynamics of protein was considered as a random walk on such
a hierarchical set and was described by the Kolmogorov--Feller equation
(master equation). Note that, mathematically, such clustering problem
is not uniquely defined. As a consequence, this leads to nonuniqueness
in the definition of the hierarchical tree that describes the set
of quasi-equilibrium states of the system. Moreover, this approach
involves purely technical difficulties in the definition of the transfer
matrix between basins of different scales. Namely, to define the elements
of the transfer matrix, one should determine the energy barriers between
local basins of states, as well as determine the number of states
in each basin. The solution of this problem is fundamentally unattainable
for real protein molecules even by the methods of modern computer
experiments. As a consequence, the approach proposed in these works
had no effective continuation. Nevertheless, the works \citep{Stillinger3,BK}
served as a basis for the development of an ultrametric approach to
the multiscale description of the conformational dynamics of protein.
The ultrametric approach is based on the representation of the basins
of quasi-equilibrium macrostates hierarchically nested into each other
by hierarchically nested balls of some ultrametric space, for which
it is convenient to take the field of $p$-adic numbers \citep{ALL,VVZ,Koch,Sh}.
This approach was developed in our previous works \citep{ABK_1999,ABKO_2002,ABO_2003,ABO_2004,AB_2008,ABZ_2009,ABZ_2011,ABZ_2013,ABZ_2014}.

The main goal of the present study is the postulative construction
of a new physical theory -- conformational dynamics of protein molecules
in the native state. As an example of the application of this theory,
we propose an analytical description of the physical experiments of
\citep{ABB,SAB} on the kinetics of CO binding to myoglobin, which,
as already mentioned, were set up for studying the properties of fluctuation
dynamic mobility of protein molecules. When constructing the theory
of conformational dynamics of protein, we apply an approach based
on $p$-adic analysis \citep{VVZ}. In this paper, we try, whenever
possible, to formulate our results and conclusions in a rigorous mathematical
form. Within the constructed theory, we propose a mathematically formalized
physical model of experiments on the kinetics of CO binding to myoglobin
and show that this model provides a complete description of the experiment
in a wide temperature range of $60\div300$ K and in an observation
time window of $10^{-7}\div10^{2}\:$ s. We also show that this theory
makes it possible to build up a unified picture of the conformational
mobility of a protein molecule and realize the fact that this mobility
changes in a self-similar way under the change of the observation
time scales. This specific feature of protein molecules, which has
remained hidden to date, significantly expands our views on the dynamic
symmetry that proteins apparently possess. It is also important that
our model provides a prediction for the behavior of the kinetic curves
of the experiment in the low-temperature range ($60\div180$ K). Namely,
the behavior of the kinetic curves should change significantly upon
the extension of the observation time window: the lower the temperature,
the wider should be the observation time window in order that the
behavior of the curves be changed. This fact can serve as a recommendation
for possible future experiments on the kinetics of CO binding to myoglobin
in an extended observation time window.

The paper is organized as follows. In Section 2, we give a brief description
of physical experiments on the study of the kinetics of CO binding
to myoglobin. Sections 2 and 3 provide an account of the ultrametric
theory of conformational dynamics of protein molecules. In Section
3, we present, at the physical level, an argumentation for the emergence
of a natural ultrametric structure on the set of conformational states
of protein. The important point in this argumentation is the definition
of the concept of ``conformational state,'' which has a slightly
different meaning compared to the traditional one adopted in biophysics.
In Section 4, we present the basic principles (postulates) of the
physical theory -- conformational dynamics of protein -- whose mathematical
formalism is the analysis over the field of $p$-adic numbers. We
well understand that this mathematics is not quite familiar to a part
of physicists and biophysicists, but it allows one to easily formalize
such a complex object as a biopolymer. Section 5 is devoted to the
solution of the Cauchy problem for the equation of $p$-adic random
walk with a reaction sink. It is the solution of this mathematical
problem that underlies our description of physical experiments on
CO binding to myoglobin at different temperatures. Section 6 is devoted
to the analysis and the physical interpretation of the solutions obtained
in Section 5 and to the comparison of the predictions of the model
with experimental results. In Appendices A, B, and C, we present the
relations and theorems that we used in Section 6 for the analysis
of the asymptotic behavior of exact solutions to the problem of $p$-adic
random walk with a reaction sink.

\section{Experiments on the kinetics of CO binding to myoglobin}

Let us give a brief overview of the experiments described in detail
in \citep{ABB,SAB}. Myoglobin protein molecules bound to CO were
irradiated by a short laser pulse. This led to photodissociation,
which breaks the bond of CO to heme iron of the active center. After
that, the kinetics of CO rebinding to heme iron was investigated on
a large time scale of $10^{-7}-10^{2}\:$ s and in a wide range of
temperatures of $60-300\mathrm{\:K}$. Here only those proteins in
which a CO molecule after photodissociation remains in the active
center (in the so-called heme pocket) were taken into consideration.
The binding kinetics for such proteins depends only on the rearrangements
of the active center, and it is this kinetics that is of primary interest.

The scheme of CO rebinding to myoglobin can be represented as follows:
\begin{equation}
Mb-CO\stackrel{h\nu}{\longrightarrow}\left[Mb^{*}\longrightarrow...\longrightarrow Mb_{1}\right]\stackrel{CO}{\longrightarrow}Mb-CO.\label{Scheme_CO_Mb}
\end{equation}
Here the symbol $Mb^{*}$ denotes the conformational states in which
protein can occur immediately after the dissociation of CO from heme
iron, and the symbol $Mb_{1}$ denotes the conformational states in
which myoglobin can be rebound to CO.

The concentration of proteins unbound to CO is a function $S\left(t,T\right)$
depending on time $t$ and temperature $T$. The curves of the concentration
of unbound myoglobin molecules as a function of time, taken from \citep{SAB},
are presented in Fig. 1.

\begin{figure*}
\includegraphics[bb=0bp 0bp 1202bp 306bp,clip,scale=0.4]{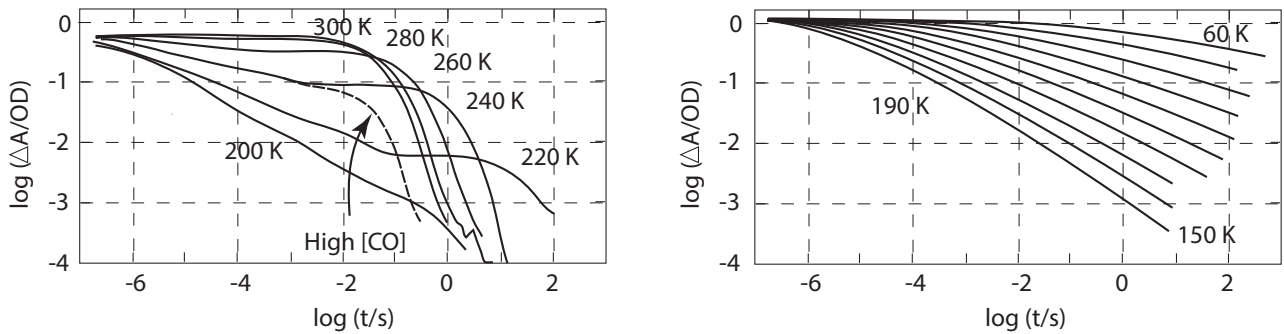}
\caption{\label{fig:wide}Concentration of unbound myoglobin molecules as a
function of time (left) in the high-temperature $200-300\mathrm{K}$
and (right) in the low-temperature $60-190\mathrm{K}$ regions \citep{SAB}.}
\end{figure*}

Depending on the behavior of $S\left(t,T\right)$, two regions are
distinguished in the temperature interval $\left(60-300\:\mathrm{K}\right)$:
the high-temperature ($200-300\:\mathrm{K}$) and the low-temperature
($60-180\:\mathrm{K}$) regions. The dependence $S\left(t,T\right)$
experimentally determined in \citep{ABB,SAB} was approximated analytically
in \citep{ZF}. In the high-temperature region, this approximation
has the form
\begin{equation}
S\left(t,T\right)=\dfrac{1}{1+\left(\dfrac{t}{\tau_{1}}\right)^{1-\tfrac{T}{T_{0}}}},\label{S(t)_exp_high}
\end{equation}
where $\tau_{1}$ is a parameter that determines the time scale. In
this temperature region, an anomalous dependence of $S\left(t,T\right)$
on temperature is observed for which the rate of the binding reaction
increases with decreasing temperature.

In the low-temperature region, the approximation of the dependence
$S\left(t,T\right)$ in the same observation time window has the form
\begin{equation}
S\left(t,T\right)=\dfrac{1}{1+\left(\dfrac{t}{\tau_{2}}\right)^{\tfrac{T}{T_{0}}}}\label{S(t)_exp_low}
\end{equation}
with a different parameter of the time scale $\tau_{2}$. This region
is characterized by normal temperature dependence, for which the rate
of the binding reaction decreases with decreasing temperature.

Let us give qualitative explanations for the curves in Fig.1. The
left panel shows the curves of the high-temperature region $\left(200-300\mathrm{\:K}\right)$.
On these curves, we can distinguish power-law and exponential regions.
At $300\mathrm{\:K}$, the binding kinetics curve is indistinguishable
from the exponential law for the given time resolution. Such behavior
is attributed to the fact that, at high temperatures, the characteristic
time of the conformational rearrangements of a myoglobin molecule
is much less than the characteristic time of CO binding to heme iron.
Therefore, the limiting time is the time of CO binding to heme iron.
As temperature decreases, the process of conformational rearrangements
of protein naturally slows down. This leads to the fact that the characteristic
times of conformational rearrangements of a myoglobin molecule increase
and become comparable with the characteristic time of CO binding to
heme iron. In this case, the kinetics of the whole process starts
to significantly depend on the evolution of the protein concentration
distribution over conformational states. This kinetics corresponds
to the power-law region, which is approximated by formula (\ref{S(t)_exp_high}).

The next thing that attracts our attention on the curves in the high-temperature
region are the regions before the exponential decay in which the concentration
of unbound molecules is almost unchanged. The reason for the presence
of such regions is purely technical: the concentration curve becomes
constant if, by a given point in time, a part of myoglobin proteins
for which the CO molecules remain in the active center after photodissociation
is exhausted. This is explained by the fact that, after photodissociation,
a CO molecule may occur in different states. In the experiment under
consideration, only two such states can be distinguished. The first
state of a CO molecule implies that the molecule is in the active
center (or the heme pocket). The second state of the CO molecule implies
that the molecule is outside the protein globule. In the latter case,
the characteristic time of repenetration of CO into the protein globule
is much greater than both the time of conformational rearrangements
of the protein molecule and the binding time of the CO molecule to
heme iron. Thus, the penetration time of CO into the globule limits
the rate of the binding reaction; therefore, the concentration of
unbound molecules is initially almost constant and then decreases
exponentially.

In the low-temperature region (Fig.1, right), the kinetic curves are
not exponential either, since in this observation time window $\left(10^{-7}-10^{2}\text{s}\right)$
we have the binding kinetics of only those CO molecules that remain
in the active center (i.e., inside the protein globule) after photodissociation.
No binding kinetics is observed for the CO molecules that remain outside
the protein after photodissociation, because the characteristic time
of penetration of a CO molecule into the body of the protein at low
temperatures is much greater than the upper boundary of the observation
time window, equal to $10^{2}\text{s}$.

The main difficulty in the description of the experiment was associated
with the stepwise change, in a rather narrow temperature interval
$\left(180-200\:\mathrm{K}\right)$, of the exponents of the power-law
approximations (\ref{S(t)_exp_high}) and (\ref{S(t)_exp_low}) at
which the temperature dependence of the binding reaction rate reverses.
The authors of \citep{ABB,SAB} themselves suggested that this change
in the kinetics is associated with the existence of a temperature
at which the protein globule passes to the glass transition phase
(see, for example, \citep{KG}), in which the behavior of protein
is qualitatively different from its behavior at room (physiological)
temperatures. In our opinion, these conclusions are incorrect for
the following reason. As already pointed out above, experiments on
the kinetics of CO binding to myoglobin were carried out in a wide
range of temperatures $\left(60-300\:\mathrm{K}\right)$ but in the
same observation time window $\left(10^{-7}-10^{2}\text{s}\right)$.
For high-temperature curves, this observation time window is rather
large and allows one to cover the whole picture of the binding kinetics.
Conversely, for low-temperature kinetics (in view of the decreased
rate of conformational rearrangements), the same time window is apparently
insufficient for observing the full picture of the kinetics of CO
binding to myoglobin. The inconsistency consists in comparing the
approximating formulas in the same time window, which leads to the
appearance of a change in the exponents in formulas (\ref{S(t)_exp_high})
and (\ref{S(t)_exp_low}) and, as a consequence, the appearance of
a stepwise change in the kinetic behavior.

\section{$p$-Adic model of conformational dynamics of protein}

In this section, we provide a physical substantiation of a $p$-adic
model for the conformational dynamics of protein and formulate the
basic principles of the model.

The configuration space $M$ of a protein molecule is a smooth manifold
defined by a set of generalized coordinates $q=\left\{ q_{i}\right\} $,
$i=1,\ldots\dim M$, corresponding to all microscopic degrees of freedom
of the molecule in the native state. Accordingly, the phase space
$P$ of a protein molecule is a smooth manifold defined by a set of
generalized coordinates and generalized momenta, $z=\left\{ q_{i},p_{i}\right\} $.
The Hamiltonian of the system $H=H\left(q,p\right)$ is a function
of its kinetic $K=K\left(q,p\right)$ and potential $U=U\left(q\right)$
energies. At a given temperature $T$ of the medium, protein executes
thermal motion, which represents a random walk on $M$ or on $P$.
The description of this random walk of protein within the Langevin
or the Fokker--Planck approaches requires a precise description of
all degrees of freedom of the protein and of the function $U\left(q\right)$,
which is hardly implementable at present. Therefore, one needs new
unconventional approaches to describe the conformational dynamics
of protein.

In 1987, Frauenfelder \citep{F} put forward the idea of ultrametricity
of proteins. Namely, to explain the experiments on the kinetics of
CO binding to a myoglobin molecule, he suggested that a protein molecule
has a set of quasi-equilibrium conformational states that are associated
with the local minima of the potential energy. He also suggested that
these conformational states can be combined into sets of states hierarchically
nested into each other and that this nesting is determined by the
value of the activation energy barrier separating any two such states.
These assumptions immediately imply that the set of quasi-equilibrium
states of protein can be mapped to the vertices of some hierarchical
three, which is what is meant by the ultrametricity of protein.

Even before Frauenfelder's paper, as well as in the first works appeared
after its publication, relatively simple models of ultrametric random
processes were proposed (see, for example, \citep{OS,HK,ZBK,KB,DS});
however, these models turned out to be hardly applicable to the description
of then available experimental data on protein dynamics. A systematic
substantiation of Frauenfelder's idea was undertaken by Becker and
Karplus in \citep{BK} (see also \citep{HS,WMW,KK,Berry,BB}).

Although these authors did not use the term ``ultrametricity,''
they provided an algorithm for constructing domains of the configuration
space, hierarchically nested into each other (which they called basins),
that correspond to quasi-equilibrium macrostates of protein, and represented
the set of all basins by the vertices of a hierarchical graph (disconnectivity
graph). A $p$-adic approach to the conformational dynamics of protein
appeared as a development of the ideas of \citep{BK,F} with the use
of the parametrization of the set of quasi-equilibrium states of protein
by subsets from the set of $p$-adic numbers. Note that the approach
of \citep{BK} was originally aimed at a precise determination of
the structure of basins by molecular dynamics methods followed by
the numerical simulation of the random walk of protein over the set
of basins. By contrast to this, the $p$-adic approach was designed
from the very beginning as an analytical theory of random walk on
an ultrametric space and was considered as an adequate approximation
to describe the conformational dynamics of protein.

According to the general idea of \citep{Stillinger3,BK}, the configuration
set $M$ can be divided (up to a set of measure zero) into subsets
-- elementary basins (or attraction basins). Each elementary basin
is associated with a local minimum of the function $U\left(q\right)$
and is defined as an open subset of points each of which satisfies
the following condition: on $M$, there exists a continuous path $q=q\left(s\right)$
from this point to a point of local minimum such that, at each point
of this path, $\dfrac{dq}{ds}=-\nabla U\left(q\right)$.

Denote by $\mathcal{B}$ the set of all elementary basins. The conformational
state (conformation) $C=C\left(B\right)$ of a protein, corresponding
to some basin $B\in\mathcal{B}$, is the process of random walk of
protein over the phase space $P$ with distribution function close
to the equilibrium distribution function in this basin,
\[
f\left(q,p\right)=\begin{cases}
Z_{B}^{-1}\exp\left(-\dfrac{H\left(q,p\right)}{kT}\right), & q\in B,\\
0, & q\notin B,
\end{cases}
\]
where $k$ is the Boltzmann constant and $Z_{B}$ is the partition
function over the subset $P$ bounded by the basin $B$. Any such
state is also called a quasistationary state of the protein in the
basin $B$. Denote the set of all conformational states by $\mathcal{U}$.

We stress that the conformational state $C\left(B\right)$ of protein
at a given time $t$ is not identified with the location of the protein
at this time at a point $q\in B$, because the conformation state
is determined by the distribution function rather than by a point
of the configuration space. The system is in some conformational state
during some random time interval starting from the time when the distribution
function of the protein becomes quasi-equilibrium in the elementary
basin $B$ and ending at time when the protein leaves the basin $B$.
The average transition time between two conformations $C\left(B^{\prime}\right)$
and $C\left(B^{\prime\prime}\right)$ corresponding to two elementary
basins $B^{\prime}$ and $B^{\prime\prime}$ is $\varDelta t=\tau\left(B^{\prime},B^{\prime\prime}\right)+\tau\left(B^{\prime\prime}\right)$,
where $\tau\left(B^{\prime},B^{\prime\prime}\right)$ is the mean
transition time from the basin $B^{\prime}$ to the basin $B^{\prime\prime}$
along some path $\varGamma\subset M$ connecting two points of the
basins $B^{\prime}$ and $B^{\prime\prime}$ and $\tau\left(B^{\prime\prime}\right)$
is the mean relaxation time to the quasi-equilibrium state in the
basin $B^{\prime\prime}$. Denote by $t\left(B\right)$ the average
residence time of the protein in the conformation $C\left(B\right)$.
Then it is natural to assume that $\tau\left(B\right)\ll t\left(B\right)$
and $\tau\left(B^{\prime},B^{\prime\prime}\right)\ll\left\{ t\left(B^{\prime}\right),t\left(B^{\prime\prime}\right)\right\} $.
The first condition follows from the very fact of the existence of
quasistationary states. The second condition follows from the following
fact. The transition path $\varGamma$ connecting two elementary basins
is unlimited; it may directly connect two basins (this is possible
if the basins have a common boundary); however, it may also pass through
other basins different from $B^{\prime}$ and $B^{\prime\prime}$.
In the first case, $\tau\left(B^{\prime},B^{\prime\prime}\right)=0$.
In the second case, one or several other intermediate basins may lie
on the transition path. Suppose that the system, during its transition
from $B^{\prime}$ into $B^{\prime\prime}$, passes through some intermediate
elementary basin $B_{int}$ and relaxes to the quasi-equilibrium state.
This is only possible if, for a given motion in the basin $B_{int}$,
the mean value of the velocity of the system on its transition path
is much greater than the rms value of its velocity in the conformational
state $C\left(B_{int}\right)$. Only in this case the system in a
short time goes outside $B_{int}$ and reaches some other basin with
high probability. In this case, the transition time through the intermediate
basin $B_{int}$ is much less than the average residence time of the
system in the state $C\left(B_{int}\right)$. In view of the aforesaid,
we assume that $\tau\left(B^{\prime},B^{\prime\prime}\right)+\tau\left(B^{\prime\prime}\right)\ll\left\{ t\left(B^{\prime}\right),t\left(B^{\prime\prime}\right)\right\} $
for any two elementary basins $B^{\prime}$ and $B^{\prime\prime}$.
This means that the dynamics of the random evolution of the protein
over the set of conformational states can be described by Markov's
stepwise random process, which takes into account only the residence
time of protein in the conformational states. In this case, the distribution
function $f_{C}\left(t\right)$ of the protein over the set of conformational
states $\mathcal{U}$ satisfies the Kolmogorov--Feller equation (master
equation) \citep{Gardiner}
\begin{equation}
\dfrac{df_{C}\left(t\right)}{dt}=\sum_{C^{\prime}\in\mathcal{U}}\left(P_{CC^{\prime}}f_{C^{\prime}}\left(t\right)-P_{C^{\prime}C}f_{C}\left(t\right)\right),\label{KF_Conf}
\end{equation}
where $P_{CC^{\prime}}$ is the matrix of transition probabilities
(in unit time) between conformations.

On the set of basins $\mathcal{B}$, we can introduce a distance function
(metric). To this end, for any two elementary basins $B^{\prime}$
and $B^{\prime\prime}$, we define a function $E\left(B^{\prime},B^{\prime\prime}\right)$
whose value is equal to the minimum of all the numbers $E$ satisfying
the following condition: there exists a path $\varGamma\subset M$
connecting the points $q_{\min}^{\prime}$ and $q_{\min}^{\prime\prime}$
of the basins $B^{\prime}$ and $B^{\prime\prime}$ such that $E=\max_{q\in\varGamma}U\left(q\right)$.
The value of the function $E\left(B^{\prime},B^{\prime\prime}\right)$
will be referred to as the value of the potential barrier between
the basins $B^{\prime}$ and $B^{\prime\prime}$. Introduce a function
\begin{equation}
d\left(B^{\prime},B^{\prime\prime}\right)=h\left(E\left(B^{\prime},B^{\prime\prime}\right)\right),\label{metr}
\end{equation}
where $h$ is an arbitrary positive increasing function. We can show
that the function (\ref{metr}) is ultrametric on the set of basins
$\mathcal{B}$ (see, for example, \citep{Koz}). Since there is one-to-one
correspondence $C=C\left(B\right)$ between conformational states
and elementary basins, the set of conformational states is also an
ultrametric space with ultrametric $d\left(C\left(B^{\prime}\right),C\left(B^{\prime\prime}\right)\right)=d\left(B^{\prime},B^{\prime\prime}\right)$.

To describe the dynamics of protein by Eq. (\ref{KF_Conf}), we need
exact parameterization of all conformations and the definition of
the matrix of transition probabilities $P_{CC^{\prime}}$. This description
requires serious simplifications of the model.

The first simplifying assumption is that $\mathcal{U}$ is assumed
to be a homogeneous ultrametric space. Recall that an ultrametric
space is homogeneous if, for any ball, the number of maximal subballs
nested into it is the same. We will call the set of conformations
the ultrametric distance between which does not exceed $r$ a ball
of radius $r$ on the space of conformations $\mathcal{U}$. As applied
to our model, this assumption implies that any ball $B_{i}\subset\mathcal{U}$
of radius $r_{i}$ is a union of an equal number $p\geq2$ of balls
$B_{i-1,a}\subset\mathcal{U}$ ($a=1,2,\ldots,p$) of radius $r_{i-1}$
nested into it; i.e., $B_{i}=\cup_{a=1}^{p}B_{i-1,a}$.

This assumption allows us to perform a $p$-adic parameterization
of the space of conformations, i.e., to map this space to the field
of $p$-adic numbers $\mathbb{Q}_{p}$. Since in $\mathbb{Q}_{p}$
every ball of radius $r_{i}=p^{i}$ is a union of $p$ subballs of
radius $r_{i-1}=p^{i-1}$ nested into it, it is natural to assign
a $p$-adic ball of given radius $r=r_{0}$ to each conformation.
Without loss of generality, we can take the value of $r_{0}$ equal
to $1$. Thus, any conformational state of protein can be parameterized
by a certain $p$-adic ball $B_{0}$ of unit radius. In this case,
any point $x\in B_{0}$ is the center of this ball and can be used
for identifying the conformation corresponding to the ball $B_{0}$.

The ultrametric distance between two conformations $C$ and $C^{\prime}$
corresponding to two $p$-adic balls $B_{0}\subset\mathbb{Q}_{p}$
and $B_{0}^{\prime}\subset\mathbb{Q}_{p}$ of radii $r_{0}=1$ is
the $p$-adic distance $d\left(x,y\right)=|x-y|_{p}$ between arbitrary
points $x\in B_{0}$ and $y\in B_{0}^{\prime}$. It is exactly equal
to
\[
|x-y|_{p}=r_{j},
\]
where $r_{j}=p^{j}$ is the radius of the minimal ball in $\mathbb{Q}_{p}$
that contains both balls $B_{0}$ and $B_{0}^{\prime}$. In this $p$-adic
parameterization of the conformational space, we will describe the
state of a protein ensemble by the distribution function $f\left(x\right)$,
$x\in\mathbb{Q}_{p}$. Here $f\left(x\right)$ is assumed to be a
locally constant function with radius of local constancy equal to
$r_{0}=1$ (i.e., for any $x\in\mathbb{Q}_{p}$ and $x'\in\mathbb{Z}_{p}$,
where $\mathbb{Z}_{p}=\left\{ x\in\mathbb{Q}_{p}:\:|x|_{p}\leq1\right\} $,
the equality $f\left(x\right)=f\left(x+x'\right)$ holds).

The following simplifying assumption concerns the matrix of transition
probabilities in unit time between conformations, $P_{CC^{\prime}}$.
We adopt that $P_{CC^{\prime}}$ is completely determined by the value
of the potential barrier between the basins $B$ and $B^{\prime}$,
which correspond to the conformations $C$ and $C^{\prime}$, i.e.,
\[
P_{CC^{\prime}}=W\left(d\left(C,C^{\prime}\right)\right)=W\left(|x-y|_{p}\right),
\]
where $W$ is a function, $x\in B_{0}$ and $y\in B_{0}^{\prime}$,
and the $p$-adic balls $B_{0}\subset\mathbb{Q}_{p}$ and $B_{0}^{\prime}\subset\mathbb{Q}_{p}$
parameterize the conformations $C$ and $C^{\prime}$, respectively.

Having adopted the above assumptions, we can formulate the following
basic principles of the $p$-adic model of the conformational dynamics
of protein.

1. The set of all possible conformational states of protein of all
levels is parameterized by a set of $p$-adic balls of unit radius
of the field of $p$-adic numbers $\mathbb{Q}_{p}$.

2. The dynamics of protein on the set of conformational states is
represented by a random walk on the field $\mathbb{Q}_{p}$, which
is described by a Markov random process $\xi\left(t,\omega\right)\colon\Omega\times\mathbb{\mathbb{R}_{+}\rightarrow\mathbb{Q}}_{p}$.
The density of the distribution function $f\left(x,t\right)$ of such
a process is assumed to be a locally constant function with radius
of constancy equal to one (i.e., for any $x$ and $x'$, $|x'|_{p}\leq1$,
the equality $f\left(x\right)=f\left(x+x'\right)$ holds), and it
is a solution of the equation of $p$-adic random walk (the Kolmogorov--Feller
equation on the field of $p$-adic numbers):
\begin{equation}
\frac{\partial f(x,t)}{dt}=\intop_{\mathbb{Q}_{p}}W\left(|x-y|_{p}\right)\left(f\left(y,t\right)-f\left(x,t\right)\right)dy.\label{EQ_URW}
\end{equation}

The following principle is necessary to reproduce the power-law relaxation
functions observed in a number of experiments in the $p$-adic model
of conformational dynamics of protein.

3. Equation (\ref{EQ_URW}) is covariant with respect to the scaling
transformations

\begin{equation}
\left\{ \begin{array}{l}
x\rightarrow x'=\lambda x,\\
t\rightarrow t'=\left|\lambda\right|_{p}^{-\alpha}t,\\
f\left(x,t\right)=f'\left(x',t'\right)=\left|\lambda\right|_{p}^{-1}f\left(\lambda x,\left|\lambda\right|_{p}^{-\alpha}t\right),
\end{array}\right.\label{cond}
\end{equation}

where $\lambda\in\mathbb{Q}_{p}$ is the transformation parameter
and $\alpha\in\mathbb{R}_{+}$.

Assumption 3 imposes a stringent constraint on the choice of the kernel
$W\left(|x-y|_{p}\right)$ of the integral operator in Eq. (\ref{EQ_URW}).
Namely, under this condition, the kernel of this operator coincides
up to a factor with the kernel of the Vladimirov operator \citep{VVZ}:
\begin{equation}
W\left(|x-y|_{p}\right)\sim\dfrac{1}{|x-y|_{p}^{\alpha+1}}.\label{W}
\end{equation}
The parameter $\alpha$ can be given a physical meaning if we set
\begin{equation}
\alpha=\dfrac{E_{0}}{kT}\label{T}
\end{equation}
and write
\begin{equation}
\dfrac{1}{|x-y|_{p}^{\alpha}}=\exp\left(-\dfrac{E_{0}\log\left(|x-y|_{p}\right)}{kT}\right),\label{Bol}
\end{equation}
where $T$ is temperature, $k$ is the Boltzmann constant, and $E_{0}$
is a parameter with the dimension of energy. In this representation,
expression (\ref{Bol}) can be interpreted as the Boltzmann factor
defining the probability that the system overcomes the potential barrier
\begin{equation}
E\left(x,y\right)=E_{0}\log|x-y|_{p}\label{E(x,y)}
\end{equation}
between two basins that correspond to a conformation containing the
points $x$ and $y$. In this case, the additional factor $\dfrac{1}{|x-y|_{p}}$
in (\ref{W}) is inversely proportional to a combinatorial factor
equal to the number of conformations whose basins are separated by
the potential barrier (\ref{E(x,y)}) from the basin of the conformation
containing the point $x$.

\section{Solution of the Cauchy problem for the equation of $p$-adic random
walk with a reaction sink}

Formally, the kinetics of CO binding to myoglobin is described by
a Cauchy problem of the form \citep{ABKO_2002,ABZ_2014}
\begin{equation}
\left\{ \begin{array}{c}
\dfrac{\partial f\left(x,t\right)}{\partial t}=-\tau^{-1}D^{\alpha}f\left(x,t\right)-\lambda\Omega\left(|x|_{p}\right)f\left(x,t\right),\\
f\left(x,0\right)=N_{r}|x|_{p}^{-\beta}\left(\Omega\left(|x|_{p}p^{-r}\right)-\Omega\left(|x|_{p}\right)\right).
\end{array}\right.\label{CP_CO_Mb}
\end{equation}
Here $\alpha>1$, $\beta>1$, $r>1$, $\lambda>0$, $\tau>0$, $D^{\alpha}$
is the Vladimirov pseudodifferential operator \citep{VVZ})
\[
D^{\alpha}\varphi\left(x\right)=\dfrac{1}{\Gamma_{p}\left(-\alpha\right)}{\displaystyle \intop_{\mathbb{Q}_{p}}\dfrac{\varphi\left(y\right)-\varphi\left(x\right)}{|x-y|_{p}^{\alpha+1}}}dy,
\]
$\Gamma_{p}\left(-\alpha\right)=\dfrac{1-p^{-\alpha-1}}{1-p^{\alpha}}$
is a $p$-adic analog of the gamma function, and the function $\Omega\left(\xi\right)$
is defined as

\[
\Omega\left(\xi\right)=\begin{cases}
1, & \xi\leq1,\\
0, & \xi>1.
\end{cases}
\]
In addition, $N_{r}=\left(\dfrac{p}{p-1}\right)\dfrac{p^{\beta-1}-1}{1-p^{-\left(\beta-1\right)r}}$
in (\ref{CP_CO_Mb}) is the normalization factor, which is determined
by the requirement ${\displaystyle \intop_{\mathbb{Q}_{p}}f\left(x,0\right)}dx=1$.
The Vladimirov operator $D^{\alpha}$ is defined on the class of functions
$W^{a},\:0\leq a<\alpha$ (i.e., on complex-valued functions $\varphi\left(x\right)$
on $\mathbb{Q}_{\text{p}}$ that satisfy the following conditions:
(1) $|\varphi\left(x\right)|\leqslant C\left(1+|x|_{p}^{a}\right)$,
$C\in\mathbb{R}_{+}$, and (2) there exists an $l\left(\varphi\right)\in\mathbb{\mathbb{N}}$
such that, for any $x\in\mathbb{\mathbb{Q}}_{p}$ and any $x'\in\mathbb{Q}_{p},\:|x'|_{p}\leqslant p^{l}$,
the equality $\varphi\left(x+x'\right)=\varphi\left(x\right)$ holds.

The physical meaning of the Cauchy problem (\ref{CP_CO_Mb}) is quite
transparent. The function $f\left(x,t\right)$ is the concentration,
normalized to unity, of protein molecules unbound to CO that are in
the conformational state parameterized by a point $x\in\mathbb{Q}_{p}$
at time $t$. We can see from the scheme (\ref{Scheme_CO_Mb}) of
CO rebinding to myoglobin that, after photodissociation, the protein
passes to the state $Mb^{*}$, which is described by the initial condition
of the Cauchy problem. Note that, based on the available experimental
data, we can say nothing about the actual distribution of protein
molecules over conformational states immediately after photodissociation;
hence, this distribution can only be a model distribution. The domain
of conformational states $Mb_{1}$ in which the reaction of CO molecule
binding to protein takes place is mathematically described by a domain
$\mathbb{Z}_{p}\subset\mathbb{Q}_{p}$ containing the reaction sink.
This sink corresponds to the term $-\lambda\Omega\left(|x|_{p}\right)f\left(x,t\right)$
in Eq. (\ref{CP_CO_Mb}) and describes a decrease in the concentration
of unbound proteins due to their binding to CO. The conformational
transitions $Mb^{*}\longrightarrow...\longrightarrow Mb_{1}$ are
described by the term with the Vladimirov pseudodifferential operator
in Eq. (\ref{CP_CO_Mb}), which is responsible for the ultrametric
diffusion of protein through conformational states. In Eq. (\ref{CP_CO_Mb}),
the parameter $\tau$ defines the time scale $t$, the parameter $\alpha$
is related to temperature by formula (\ref{T}), the parameter $\lambda$
is the rate of CO binding to myoglobin in unit time, and the parameter
$\beta$ characterizes the initial distribution of unbound myoglobin
over conformations. To match the model with experiment, we impose
the requirements $\alpha>1$ and $\beta>1$. In theoretical calculations,
we set $\tau=1;$ thus, $\lambda,\:t,\:x,$ and $f$ are dimensionless
parameters.

\textbf{Theorem 1.} \textit{A solution of the Cauchy problem (\ref{CP_CO_Mb})
in the class of $f\left(x,t\right)\in W^{a}\cap L^{1}\left(\mathbb{Q}_{p},dx\right)\cap C^{1}\left(\mathbb{R}_{+}\right)$
exists and is unique. }

\textbf{Proof.} The Cauchy problem (\ref{CP_CO_Mb}) in terms of Fourier
transforms has the form \begin{widetext}
\begin{equation}
\left\{ \begin{array}{c}
\dfrac{\partial}{\partial t}\tilde{f}\left(k,t\right)=-|k|_{p}^{\alpha}\tilde{f}\left(k,t\right)-\lambda\intop_{Q_{p}}\Omega\left(|k-q|_{p}\right)\tilde{f}\left(q,t\right)dq,\\
\tilde{f}\left(k,0\right)=N_{r}\left(\dfrac{\left(1-p^{-1}\right)}{p^{\beta-1}-1}\left(\Omega\left(|k|_{p}\right)-p^{-\left(\beta-1\right)r}\Omega\left(|k|_{p}p^{r}\right)\right)\right.\left.-\dfrac{1-p^{-\beta}}{p^{\beta-1}-1}|k|_{p}^{\beta-1}\left(\Omega\left(|k|_{p}\right)-\Omega\left(|k|_{p}p^{r}\right)\right)\right)
\end{array}\right.\label{F_sys}
\end{equation}
\end{widetext}

If $|k|_{p}>1,$ then $\tilde{f}\left(k,t\right)\equiv0;$ this follows
from the Fourier transform of the initial condition $\tilde{f}\left(k,0\right)=0,\:|k|_{p}>1.$
If $|k|_{p}\leq1$, then we have
\begin{equation}
\dfrac{\partial}{\partial t}\widetilde{f}\left(k,t\right)=-|k|_{p}^{\alpha}\widetilde{f}\left(k,t\right)-\lambda{\displaystyle \intop_{Z_{p}}}\tilde{f}\left(q,t\right)dq.\label{F_1}
\end{equation}
Let us show that there exists a Laplace transform for the function
$\tilde{f}\left(k,t\right)$. In view of the inequality
\[
\left|\widetilde{f}\left(k,t\right)\right|\leq{\displaystyle \intop_{\mathbb{Q}_{p}}\left|f\left(x,t\right)\right|dx}<\infty,
\]
the function $\widetilde{f}\left(k,t\right)$ is bounded on $\mathbb{Z}_{p}$;
moreover, it is continuous with respect to the variable $k$. With
respect to the variable $t\in\mathbb{R}_{+}$, the function $\widetilde{f}\left(k,t\right)$
is continuous and differentiable. Let us integrate Eq. (\ref{F_1})
over $\mathbb{Z}_{p}$. Then we have
\[
\dfrac{\partial}{\partial t}{\displaystyle {\displaystyle \intop_{Z_{p}}\tilde{f}\left(k,t\right)dk}=-\intop_{Z_{p}}|k|_{p}^{\alpha}\tilde{f}\left(k,t\right)dk-\lambda\intop_{Z_{p}}\tilde{f}\left(q,t\right)dq},
\]
which implies \begin{widetext}
\[
\left|\dfrac{\partial}{\partial t}\intop_{Z_{p}}\tilde{f}\left(k,t\right)dk\right|=\left|\intop_{\mathbb{Z}_{p}}|k|_{p}^{\alpha}\tilde{f}\left(k,t\right)dk+\lambda\int_{\mathbb{Z}_{p}}\tilde{f}\left(q,t\right)dq\right|\leqslant\left|\intop_{Z_{p}}|k|_{p}^{\alpha}\tilde{f}\left(k,t\right)dk\right|+\left|\lambda\intop_{Z_{p}}\tilde{f}\left(q,t\right)dq\right|.
\]
\end{widetext}

Denoting $S_{\mathbb{Z}_{p}}\left(t\right)={\displaystyle \intop_{\mathbb{Z}_{p}}\tilde{f}\left(k,t\right)dk}$,
we write $\left|\dfrac{\partial}{\partial t}S_{\mathbb{Z}_{p}}\left(t\right)\right|\leqslant\left|\left(1+\lambda\right)S_{\mathbb{Z}_{p}}\left(t\right)\right|$
or $-\left(1+\lambda\right)dt\leqslant\dfrac{dS_{\mathbb{Z}_{p}}\left(t\right)}{S_{\mathbb{Z}_{p}}\left(t\right)}\leqslant\left(1+\lambda\right)dt,$
whence we obtain $S_{\mathbb{Z}_{p}}\left(t\right)\leqslant A\exp\left(\left(1+\lambda\right)t\right)$
for some $A$. From the last inequality we obtain ${\displaystyle \intop_{\mathbb{Z}_{p}}\tilde{f}\left(k,t\right)dk\leqslant A\exp\left[\left(1+\lambda\right)t\right]}$,
which implies that $\tilde{f}\left(k,t\right)\leqslant M\exp\left[s_{0}t\right]$.
This upper bound of the function $\tilde{f}\left(k,t\right)$ proves
that, for this function, there exists a Laplace transform, which we
denote by $\widetilde{F}\left(k,s\right)$.

In terms of $\widetilde{F}\left(k,s\right)$, the Cauchy problem (\ref{CP_CO_Mb})
is rewritten as
\[
s\widetilde{F}\left(k,s\right)=\tilde{f}\left(k,0\right)-|k|_{p}^{\alpha}\widetilde{F}\left(k,s\right)-\lambda\intop_{\mathbb{Z}_{p}}\tilde{F}\left(q,s\right)dq,
\]
whence we have
\begin{equation}
\tilde{F}\left(k,s\right)=\dfrac{\tilde{f}\left(k,0\right)}{s+|k|_{p}^{\alpha}}-\lambda\dfrac{1}{s+|k|_{p}^{\alpha}}G\left(s\right),\label{F(k,s)}
\end{equation}
where
\begin{equation}
G\left(s\right)=\intop_{\mathbb{Z}_{p}}\tilde{F}\left(q,s\right)dq.\label{G(s)}
\end{equation}
Integrating (\ref{F(k,s)}) with respect to $k\in Z_{p}$, we obtain

\[
G\left(s\right)=\intop_{Z_{p}}\dfrac{\tilde{f}\left(\left(k,0\right)\right)}{s+|k|_{p}^{\alpha}}dk-\lambda\intop_{Z_{p}}\dfrac{dk}{s+|k|_{p}^{\alpha}}G\left(s\right);
\]
hence,
\begin{equation}
G\left(s\right)=\dfrac{\intop_{Z_{p}}\dfrac{\tilde{f}\left(\left(k,0\right)\right)}{s+|k|_{p}^{\alpha}}dk}{1+\lambda\intop_{Z_{p}}\dfrac{dk}{s+|k|_{p}^{\alpha}}}.\label{G(s)_int}
\end{equation}
The calculation of the integrals in (\ref{G(s)_int}) yields
\begin{equation}
G\left(s\right)=\dfrac{J\left(s\right)+h_{r}\left(s\right)-H_{r}\left(s\right)}{1+\lambda J\left(s\right)},\label{G(s)_frac}
\end{equation}
where
\[
J\left(s\right)=\left(1-p^{-1}\right){\displaystyle \sum_{n=0}^{\infty}\dfrac{p^{-n}}{s+p^{-\alpha n}}},
\]

\[
h_{r}\left(s\right)=\left(1-p^{-1}\right)\dfrac{p^{-\left(\beta-1\right)r}}{1-p^{-\left(\beta-1\right)r}}{\displaystyle \sum_{n=0}^{r-1}\dfrac{p^{-n}}{s+p^{-\alpha n}},}
\]

\[
H_{r}\left(s\right)=\dfrac{1-p^{-\beta}}{1-p^{-\left(\beta-1\right)r}}{\displaystyle \sum_{n=0}^{r-1}\dfrac{p^{-\beta n}}{s+p^{-\alpha n}}.}
\]
Consider the function $G\left(s\right)$. In the domain $\mathrm{Re}s>0$,
it is holomorphic. In what follows, it is convenient to consider this
function on the extended complex plane. To this end, we define it
at removable points $s=-p^{-\alpha k},\;k=0,1,2,...$ , where it is
not defined but has finite limits
\[
\lim_{s\rightarrow-p^{-\alpha k}}G\left(s\right)=
\]
\[
\dfrac{1}{\lambda}\left(1+\dfrac{p^{-\left(\beta-1\right)r}}{1-p^{-\left(\beta-1\right)r}}-\dfrac{1-p^{-\beta}}{1-p^{-\left(\beta-1\right)r}}\dfrac{p}{p-1}p^{\left(\beta-1\right)k}\right)
\]
for $k\leqslant r-1$ and
\[
\lim_{s\rightarrow-p^{-\alpha k}}G\left(s\right)=\dfrac{1}{\lambda}
\]
for $k>r-1.$ Then the function (\ref{G(s)_frac}) is holomorphic
on the entire extended complex plane except for the points $s=-\lambda_{k},\;k=-1,0,1,2,...$,
where it has simple poles determined from the equation
\begin{equation}
1+\lambda J\left(s\right)=0.\label{eq_pol}
\end{equation}
It is easily seen that the values $\lambda_{k}$ satisfy the inequality
\[
p^{-\alpha\left(k+1\right)}<\lambda_{k}<p^{-\alpha k},\:k=0,1,2,\ldots,\:\lambda_{-1}>1.
\]
The function $G\left(s\right)$ is not meromorphic since $s=0$ is
an essentially singular point at which the poles are condensed. Notice
that ${\displaystyle \lim_{s\rightarrow0,\:\mathrm{Re}s>0}G\left(s\right)=\dfrac{1}{\lambda}}$
and ${\displaystyle \lim_{s\rightarrow\infty}G\left(s\right)=0}$
uniformly with respect to $\arg s$. Let us change the variable: $s\rightarrow z=\dfrac{1}{s}$.
Then the auxiliary function $G\left(\dfrac{1}{z}\right)=\Phi\left(z\right)$
is meromorphic; moreover, ${\displaystyle \lim_{z\rightarrow\infty}\Phi\left(z\right)=\dfrac{1}{\lambda}}$
and ${\displaystyle \lim_{z\rightarrow0}\Phi\left(z\right)=0}$. Since
$|\Phi\left(z\right)|\leq A|z|^{m}$ for $z\rightarrow\infty$, Mittag--Leffler's
simple pole expansion theorem implies that $\Phi\left(z\right)$ can
be represented as a uniformly convergent (except for a countable number
of simple poles) series $\Phi\left(z\right)={\displaystyle \sum_{k=-1}^{\infty}\dfrac{a_{k}}{z+\frac{1}{\lambda_{k}}}}$.
Thus, the function $G\left(s\right)=\Phi\left(\dfrac{1}{s}\right)$
can also be represented as a uniformly convergent series
\begin{equation}
G\left(s\right)={\displaystyle \sum_{k=-1}^{\infty}\dfrac{b_{k}}{s+\lambda_{k}},}\label{G_poles}
\end{equation}
where $b_{k}$ are the residues of the function $G\left(s\right)$
at the poles $s=-\lambda_{k}$, given by

\begin{equation}
b_{k}={\displaystyle \underset{s=-\lambda_{k}}{\mathrm{Res}G\left(s\right)}}=\dfrac{1}{\lambda^{2}|J^{\prime}\left(-\lambda_{k}\right)|}+\dfrac{h_{r}\left(-\lambda_{k}\right)-H_{r}\left(-\lambda_{k}\right)}{-\lambda|J^{\prime}\left(-\lambda_{k}\right)|}.\label{b_k}
\end{equation}
Thus, the solution in terms of Fourier transforms has the form
\[
\tilde{f}\left(k,t\right)=\tilde{f}\left(k,0\right)\exp\left(-|k|_{p}^{\alpha}t\right)-
\]

\begin{equation}
\lambda\exp\left(-|k|_{p}^{\alpha}t\right){\displaystyle \sum_{n=-1}^{\infty}\dfrac{b_{n}}{\lambda_{n}-|k|_{p}^{\alpha}}\left(1-\exp\left[-\left(\lambda_{n}-|k|_{p}^{\alpha}\right)t\right]\right).}\label{f(k,t)}
\end{equation}
We can show by direct substitution of (\ref{f(k,t)}) into Eq. (\ref{F_sys})
that this solution is a solution of the Cauchy problem (\ref{F_sys})
in terms of Fourier transforms. Similarly we can show that the function

\[
f\left(x,t\right)={\displaystyle \intop_{\mathbb{Q}_{p}}\tilde{f}\left(k,t\right)\chi\left(-kx\right)dx}
\]
satisfies Eq. (\ref{CP_CO_Mb}), i.e., that a solution exists. The
coefficients $b_{n}$ are determined uniquely for a given initial
condition; this implies the uniqueness of the solution. The theorem
is proved.

\section{Analysis of the solutions and relation to experiment}

The concentration $S\left(t\right)$ of myoglobin molecules that are
not bound to CO at time $t$ is
\[
S\left(t\right)=\intop_{\mathbb{Q}_{p}}f\left(x,t\right)dx.
\]
If we integrate Eq. (\ref{CP_CO_Mb}) with respect to $\mathbb{Q}_{p}$,
we arrive at the equation
\begin{equation}
\dfrac{\partial S\left(t\right)}{\partial t}=-\lambda S_{Z_{p}}\left(t\right),\label{dS(t)/dt}
\end{equation}
where
\[
S_{Z_{p}}\left(t\right)=\intop_{\mathbb{Z}_{p}}f\left(x,t\right)dx
\]
is the concentration of proteins in the conformational states $Mb_{1}$
and $\lambda$ is the rate of CO binding to myoglobin. The function
$S_{Z_{p}}\left(t\right)$ is the Laplace transform of the meromorphic
function $G\left(s\right)$; it can be represented as an infinite
series:
\begin{equation}
S_{Z_{p}}\left(t\right)={\displaystyle \sum_{k=-1}^{\infty}}b_{k}\exp\left(-\lambda_{k}t\right)\risingdotseq G\left(s\right)={\displaystyle \sum_{k=-1}^{\infty}\dfrac{b_{k}}{s+\lambda_{k}},}\label{S_Z_p}
\end{equation}
where $\lambda_{k}$ and $b_{k}$ are, respectively, the poles and
residues of $G\left(s\right)$. Then, taking into account that $G\left(0\right)=\dfrac{1}{\lambda}$,
from (\ref{dS(t)/dt}) and (\ref{S_Z_p}) we obtain
\begin{equation}
S\left(t\right)=\lambda\sum_{k=-1}^{\infty}\dfrac{b_{k}}{\lambda_{k}}\exp\left(-\lambda_{k}t\right).\label{S(t)_sum}
\end{equation}
Let us represent (\ref{S(t)_sum}) as
\[
S\left(t\right)=\lambda\dfrac{b_{-1}}{\lambda_{-1}}\exp\left(-\lambda_{-1}t\right)+\widetilde{S}\left(t\right),
\]
where
\begin{equation}
\widetilde{S}\left(t\right)=\lambda\sum_{k=0}^{\infty}\dfrac{b_{k}}{\lambda_{k}}\exp\left(-\lambda_{k}t\right).\label{tilde_S}
\end{equation}
Taking into account (\ref{App_A_Result}) and (\ref{b_1})--(\ref{b_3_(3)}),
we write
\[
S_{1}\left(t\right)<\widetilde{S}\left(t\right)<S_{2}\left(t\right),
\]
where $S_{1}\left(t\right)$ and $S_{2}\left(t\right)$ have the following
structure: \begin{widetext}
\[
S_{1}\left(t\right)=A_{1}{\displaystyle \sum_{k=0}^{\infty}}p^{-\left(\alpha-1\right)k}\exp\left(-p^{-\alpha k}t\right)+B_{1}{\displaystyle \sum_{k=0}^{r-1}}p^{-\left(\beta-1\right)k}\exp\left(-p^{-\alpha k}t\right)+C_{1}{\displaystyle \left(p^{\left(\alpha-\beta\right)r}-1\right)\sum_{k=r}^{\infty}}p^{-\left(\alpha-1\right)k}\exp\left(-p^{-\alpha k}t\right)
\]
\begin{equation}
-D_{1}p^{-\left(\beta-1\right)r}{\displaystyle \sum_{k=0}^{r-1}\exp\left(-p^{-\alpha k}t\right)}-E_{1}{\displaystyle \left(p^{\left(\alpha-1\right)r}-1\right)\sum_{k=r}^{\infty}}p^{-\left(\alpha-1\right)k}{\displaystyle \exp\left(-p^{-\alpha k}t\right)},\label{S_1}
\end{equation}

\[
S_{2}\left(t\right)=A_{2}{\displaystyle \sum_{k=0}^{\infty}}p^{-\left(\alpha-1\right)k}\exp\left(-p^{-\alpha k}p^{-\alpha}t\right)
\]
\[
+B_{2}{\displaystyle \sum_{k=0}^{r-1}}p^{-\left(\beta-1\right)k}\exp\left(-p^{-\alpha k}p^{-\alpha}t\right)+C_{2}{\displaystyle p^{\alpha r}\left(1-p^{-\beta r}\right)\sum_{k=r}^{\infty}}p^{-\left(\alpha-1\right)k}\exp\left(-p^{-\alpha k}p^{-\alpha}t\right)
\]
\begin{equation}
-D_{2}p^{-\left(\beta-1\right)r}{\displaystyle \sum_{k=0}^{r-1}p^{-\left(\alpha-1\right)k}\exp\left(-p^{-\alpha k}p^{-\alpha}t\right)}-E_{2}p^{\alpha r}\dfrac{1-p^{-\beta r}}{p^{\left(\beta-1\right)r}}{\displaystyle \sum_{k=r}^{\infty}p^{-\left(\alpha-1\right)k}\exp\left(-p^{-\alpha k}p^{-\alpha}t\right)}.\label{S_2}
\end{equation}
\end{widetext}

In (\ref{S_1}) and (\ref{S_2}), the coefficients $A_{i}$, $B_{i}$,$C_{i}$,
$D_{i}$, and $E_{i}$ depend only on the parameters $\lambda$, $\alpha$,
and $\beta$, and their explicit form is unimportant for our further
analysis.

Let us find asymptotic estimates for (\ref{S_1}) and (\ref{S_2})
in the high-temperature and low-temperature regimes. In our model,
we assume that all values of the temperature parameter $\alpha$ that
satisfy the following condition correspond to the high-temperature
region:
\begin{equation}
\alpha\text{<}\beta.\label{alpha<beta}
\end{equation}
Accordingly, the low-temperature region is described by the inequality
\begin{equation}
\alpha>\beta.\label{alpha>beta}
\end{equation}
In the high-temperature region (\ref{alpha<beta}), we can consider
two cases. In case 1, we deal with large observation times such that
$t\gg p^{\beta r}$. In case 2, we deal with intermediate observation
times such that $1\ll t\ll p^{\beta r}$.

Consider case 1 and find the asymptotic behavior of (\ref{S_1}) and
(\ref{S_2}) for a fixed $r$ as $t\rightarrow\infty$. Using Theorem
1 in Appendix C, we find that the main contribution to the asymptotic
behavior of (\ref{S_1}) and (\ref{S_2}) is given only by the sums
multiplying the coefficients $A_{i}$, $C_{i}$, and $E_{i}$, which
have the following asymptotic estimates as $t\rightarrow\infty$:
\begin{equation}
S_{1}\left(t\right)>a_{1}t^{-\tfrac{\alpha-1}{\alpha}}\left(1+\omega\left(t\right)\right),\label{S_1r_high_T}
\end{equation}

\begin{equation}
S_{2}\left(t\right)<a_{2}t^{-\tfrac{\alpha-1}{\alpha}}\left(1+\omega\left(t\right)\right),\label{S_2r_high_T}
\end{equation}
where $a_{i}$ are some coefficients independent of $t$ and the symbol
$\omega\left(t\right)$ denotes functions, not explicitly shown, that
are infinitesimal as $t\rightarrow\infty$.

In case 2, we should consider the behavior of (\ref{S_1}) and (\ref{S_2})
as $t\rightarrow\infty$ and $r\rightarrow\infty$ under the condition
that $p^{-\beta r}t\rightarrow0$. In this case, the terms with coefficients
$C_{i}$, $D_{i}$, and $E_{i}$ do not contribute in the limit as
$r\rightarrow\infty$. From Theorems 1 and 2 in Appendix C, for the
sums multiplying the coefficients $A_{i}$ and $B_{i}$, respectively,
we find that, for $\alpha<\beta$, a contribution to the asymptotic
estimates $S_{1}\left(t\right)$ and $S_{2}\left(t\right)$ is made
only by the sums multiplying the coefficients $A_{i}$. In this case,
we again obtain asymptotic estimates in the form (\ref{S_1r_high_T})
and (\ref{S_2r_high_T}) but with different coefficients $a_{i}$.

Thus, for $\alpha<\beta$, we obtain estimates (\ref{S_1r_high_T})
and (\ref{S_2r_high_T}) both for intermediate observation times ($1\ll t\ll p^{\beta r}$)
and for large observation times ($t\gg p^{\beta r}$); these estimates
completely agree with formula (\ref{S(t)_exp_high}), which approximates
the experimental dependence in the high-temperature ($T\sim190-300\:\mathrm{K}$)
region:
\begin{equation}
S\left(t\right)=\dfrac{1}{1+\left(\dfrac{t}{\tau_{1}}\right)^{1-\tfrac{T}{T_{0}}}}\sim\left(\dfrac{t}{\tau_{1}}\right)^{\tfrac{T}{T_{0}}-1}=\left(\dfrac{t}{\tau_{1}}\right)^{-\tfrac{\alpha-1}{\alpha}}.\label{S_high_T}
\end{equation}
This means that the behavior of $S\left(t\right)$ in the high-temperature
region is universal and does not depend on the observation time window.
Our result shows that the behavior of $S\left(t\right)$ is independent
of the form of the initial condition, i.e., of the parameter $\beta$,
which parameterizes the density of the distribution function of proteins
over conformations after photodissociation. This is explained by the
fact that the random walk at high temperatures is rather intense.
Hence, within relatively small times compared with the observation
time ($10^{2}$ s), the concentration of unbound molecules, first,
is uniformly distributed over the domain $B_{r}$, and then is distributed
over $\mathbb{Q}_{p}\setminus B_{r}$. It is the random walk over
the domain $\mathbb{Q}_{p}\setminus B_{r}$, which is sufficiently
far from the domain $Z_{p}$ and contains the sink, that determines
the law (\ref{S_high_T}). The anomalous dependence of the binding
reaction rate on temperature (i.e., the increase in $\dfrac{dS\left(t\right)}{dt}$
with decreasing temperature) is explained equally easily. Indeed,
the higher the temperature, the further goes the random trajectory
from the support of the initial distribution in the space $\mathbb{Q}_{p}$,
and the more time it takes to reach the sink region.

In the low-temperature region, we have (\ref{alpha>beta}). Here we
can also consider two cases: case 1 -- large observation times (or
$t\gg p^{\beta r}$) and case 2 -- intermediate observation times
(or $1\ll t\ll p^{\beta r}$).

In case 1, repeating precisely the arguments for case 1 in the high-temperature
region (\ref{alpha<beta}), we obtain the same result (\ref{S_1r_high_T})
and (\ref{S_2r_high_T}) from (\ref{S_1}) and (\ref{S_2}). Nevertheless,
the result in case 2 will be different. It is this case that corresponds
to observation times of $10^{-7}\div10^{2}$ s in the experiment at
low temperatures ($T\sim60-180\:\mathrm{K}$). Indeed, consider (\ref{S_1})
and (\ref{S_2}) as $t\rightarrow\infty$, $r\rightarrow\infty$ under
the condition $p^{-\beta r}t\rightarrow0$. In this case, the terms
with the coefficients $C_{i}$, $D_{i}$, and $E_{i}$ do not contribute
in the limit as $r\rightarrow\infty$. Using Theorems 1 and 2 in Appendix
C for the sums multiplying the coefficients $A_{i}$ and $B_{i}$,
we find that, for $\alpha>\beta$, only the sums multiplying the coefficients
$B_{i}$ contribute to the asymptotic estimates for $S_{1}\left(t\right)$
and $S_{2}\left(t\right)$. As a result, we obtain
\begin{equation}
S_{1}\left(t\right)>a_{1}t^{-\tfrac{\beta-1}{\alpha}}\left(1+\omega\left(t\right)\right),\label{S_1r_low_T}
\end{equation}

\begin{equation}
S_{2}\left(t\right)<a_{2}t^{-\tfrac{\beta-1}{\alpha}}\left(1+\omega\left(t\right)\right).\label{S_2r_low_T}
\end{equation}
We can see that, in the low-temperature region for intermediate observation
times $1\ll t\ll p^{\beta r}$, the asymptotic behavior of $S\left(t\right)$
significantly depends on the parameter $\beta$, which appears in
the initial distribution. Note that (\ref{S_1r_low_T}) and (\ref{S_2r_low_T})
agree with the approximating function (\ref{S(t)_exp_low}) if we
set $\beta=2$:
\begin{equation}
S\left(t\right)=\dfrac{1}{1+\left(\dfrac{t}{\tau_{2}}\right)^{\tfrac{T}{T_{0}}}}\sim\left(\dfrac{t}{\tau_{2}}\right)^{-\tfrac{T}{T_{0}}}=\left(\dfrac{t}{\tau_{2}}\right)^{-\tfrac{1}{\alpha}}\label{S_low_T}
\end{equation}
.

\section{Discussion}

We have shown that the model considered reproduces the asymptotic
behavior of functions approximating the dependence of the concentrations
of unbound molecules in the experiments on CO binding to myoglobin.
In spite of the fact that the functions approximating the experimental
curves in the high-temperature ($T\sim190-300\:\mathrm{K}$) and low-temperature
($T\sim60-180\:\mathrm{K}$) regions have different forms, the time
dependence of $S\left(t\right)$ for unbound molecules is described
in our model by a universal function in the entire range of temperatures
from $60\:K$ to $300\:K$.

The most important consequence of our description of the behavior
of $S\left(t\right)$ in the low-temperature region is that, unpon
the extension of the observation time window, the behavior of $S\left(t\right)$
is changed. This follows from the fact that, for sufficiently large
times $t$ in the low-temperature region (\ref{alpha>beta}), we pass
to case 1 ( $t\gg p^{\beta r}$, see Section 6). Therefore, for increasing
observation times in the low-temperature region, the temperature dependence
of $S\left(t\right)$ should change, and, instead of the dependence
(\ref{S_low_T}), we should observe the same dependence (\ref{S_high_T})
as that for the high-temperature region. This is the main nontrivial
prediction of our theory, which can immediately be checked in possible
future experiments on CO binding to myoglobin in extended observation
time windows of $>100$ s, which were not covered in the experiments
of \citep{ABB,SAB}.

An interesting question is that why precisely the choice of the initial
condition in the form
\begin{equation}
f\left(x,0\right)\sim|x|_{p}^{-\beta}\left(\Omega\left(|x|_{p}p^{-r}\right)-\Omega\left(|x|_{p}\right)\right)\label{i.e.}
\end{equation}
leads to agreement with experiment for $\beta=2$ in the low-temperature
region.

The experiments carried out do not allow us to make any conclusions
about the distribution of unbound protein molecules over conformational
states immediately after photodissociation. It is known that, before
the experiment, the test sample was kept for a rather long time at
a certain temperature $\left(T\sim300\:\mathrm{K}\right)$. In this
case, the equilibrium is reached in the protein ensemble. After that,
at time $t_{0}$, the temperature is reduced to some value $T_{1}<T$.
Starting from time $t_{0}$ until the time of photodissociation $t_{1}$,
the protein ensemble does not yet have time to completely reach the
equilibrium state. At time $t_{2}$ after photodissociation (the interval
$t_{2}-t_{1}$ is the same for all experiments with different temperatures
$T_{1}$), the concentration of unbound protein molecules is measured.
The initial condition (\ref{i.e.}) is the distribution of unbound
protein molecules at time $t_{2}$. Naturally, it is impossible to
explain the origin of this initial condition within the present model.
Nevertheless, we can go beyond the model and try to imagine some hypothetical
scenario that would justify to some extent the choice of the initial
condition (\ref{i.e.}).

We can assume that, by the time $t_{0}$ when the temperature is reduced
to $T_{1}$, the nonequilibrium distribution of protein over conformational
states is described by the distribution function with support in the
neighborhood of $\mathbb{Z}_{p}$. Suppose also that the variation
of the distribution function is described by the equation of $p$-adic
random walk $\dfrac{\partial f\left(x,t\right)}{\partial t}=-\tau^{-1}D^{\alpha}f\left(x,t\right)$
and that photodissociation hardly changes the distribution of proteins
over conformational states. In this case, the distribution of unbound
molecules at time $t_{2}$ should be a solution of the Cauchy problem
for the given equation with the initial condition on a compact set.
It is known that the solution of this Cauchy problem is estimated
by a function that has asymptotics $\dfrac{C}{|x|_{p}^{\alpha+1}}t$
as $\dfrac{t}{|x|_{p}^{\alpha}}\rightarrow0$ (see, for example, \citep{Koch,ABO_2003}).
If $t=t_{2}-t_{0}$ is time passed from the beginning of cooling of
the sample to temperature $T_{1}$ to the moment of observation, then,
for small $t$, the main part of molecules are located near $\mathbb{Z}_{p}$
and are rapidly bound to CO molecules. The remaining small part of
protein molecules that are not bound to CO, which are the object of
observation in the experiment, are distributed with respect to $\mathbb{Q}_{p}$
by the law $\dfrac{Ct_{2}}{|x|_{p}^{\alpha+1}}$, where $\alpha=\dfrac{T_{0}}{T}$
is a temperature parameter (see (\ref{T})). The value $T_{0}$ of
the temperature scale corresponds to the maximum temperature of the
experiments, which coincides with the preparation temperature of the
samples $T_{0}\sim300\:\mathrm{K}$. For $T_{0}=300\:\mathrm{K}$,
the value of $T=300\:\mathrm{K}$ corresponds to the value $\alpha=1$.
Therefore, the initial distribution should have the form $\sim|x|_{p}^{-2}$,
which means $\beta=2$.
\begin{acknowledgments}
The study was supported in part by the Ministry of Education and Science
of Russia by State assignment to educational and research institutions
under project FSSS-2020-0014.
\end{acknowledgments}

\section*{Data Availability Statement}

The data supporting the findings of this study are available within
the article and its its supplementary material. All other relevant
source data are available from the corresponding author upon reasonable
request.


\appendix

\section{Estimate of $\lambda_{k}$. }

\label{A_A} Taking into account that ${\displaystyle \lim_{\lambda\rightarrow0}}\lambda_{k}=p^{-\alpha\left(k+1\right)}$,
it is convenient to use the following representation for $\lambda_{k}$:

\begin{equation}
\lambda_{k}=p^{-\alpha\left(k+1\right)}+p^{-\alpha k}\left(1-p^{-\alpha}\right)\delta_{k}.\label{lambda_k_delta_k}
\end{equation}
From the graphical solution of the equation $1+\lambda J\left(s\right)=0$,
we can show that $0<\delta_{k}<1$. Then (\ref{lambda_k_delta_k})
implies the following estimate for $\lambda_{k}$:

\[
p^{-\alpha k}p^{-\alpha}<\lambda_{k}<p^{-\alpha k},
\]
From the equation $1+\lambda J\left(-\lambda_{k}\right)=0$ we have

\begin{equation}
{\displaystyle \sum_{n=0}^{\infty}}\dfrac{p^{-n}}{p^{-\alpha n}-\lambda_{k}}=-\dfrac{p}{p-1}\dfrac{1}{\lambda}.\label{eq_lambda_k}
\end{equation}
Singling out the $k+1$-th term in the sum on the left-hand side of
Eq. (\ref{eq_lambda_k}), multiplying it by $p^{k}$, dividing by
$p^{\alpha k}$, and taking into account (\ref{lambda_k_delta_k}),
we can write (\ref{eq_lambda_k}) in the form:
\[
\dfrac{p^{-1}}{\Delta_{k}}=\dfrac{p}{p-1}\dfrac{p^{-\left(\alpha-1\right)k}}{\lambda}+
\]

\begin{equation}
{\displaystyle \sum_{i=0}^{k}\dfrac{p^{i}}{p^{\alpha i}-p^{-\alpha}-\Delta_{k}}{\displaystyle -\sum_{j=2}^{\infty}\dfrac{p^{-j}}{p^{-\alpha}-p^{-\alpha j}+\Delta_{k}}}}\label{eq_lambda_k_frac}
\end{equation}
where $\Delta_{k}=\left(1-p^{-\alpha}\right)\delta_{k}$ .

To obtain an upper bound for $\delta_{k}$, we write the inequality

\begin{equation}
\dfrac{p^{-1}}{\Delta_{k}}\text{>}{\displaystyle \dfrac{1}{1-p^{-\alpha}-\varDelta_{k}}{\displaystyle -\dfrac{p^{-1}}{p-1}\dfrac{1}{\varDelta_{k}}}},\label{B_neq_1}
\end{equation}
which follows from (\ref{eq_lambda_k_frac}). After straightforward
transformations, we obtain the following expression from (\ref{B_neq_1}):

\[
\delta_{k}<p^{-1}.
\]
To obtain a lower bound for $\delta_{k}$, we write the inequality

\begin{equation}
\dfrac{p^{-1}}{\Delta_{k}}<\dfrac{p}{p-1}\dfrac{p^{-\left(\alpha-1\right)k}}{\lambda}+{\displaystyle {\displaystyle \dfrac{1}{1-p^{\alpha}-\Delta_{k}}\dfrac{p^{\left(\alpha-1\right)}}{p^{\left(\alpha-1\right)}-1}}},\label{B_neq_2}
\end{equation}
which also follows from (\ref{eq_lambda_k_frac}). From (\ref{B_neq_2})
we obtain
\begin{equation}
d\delta_{k}^{2}-\left(1+b_{\alpha}+d_{k}\right)\delta_{k}+1<0,\label{B_neq_3}
\end{equation}
where $d_{k}=\dfrac{p^{2}}{p-1}\dfrac{p^{\alpha}-1}{p^{\alpha}}\dfrac{p^{-\left(\alpha-1\right)k}}{\lambda}$
and ${\displaystyle {\displaystyle b_{\alpha}=\dfrac{p^{\alpha}}{p^{\left(\alpha-1\right)}-1}}}$.
Solving (\ref{B_neq_3}) and performing straightforward transformations,
we obtain
\[
\delta_{k}<\dfrac{1-p^{-\alpha+1}}{1+p}.
\]
Finally, we write
\begin{equation}
\dfrac{1-p^{-\alpha+1}}{1+p}=\delta_{\min}<\delta_{k}<\delta_{\max}=p^{-1}.\label{App_A_Result}
\end{equation}

\section{Estimate of $b_{k}$. }

\label{A_B}

Let us rewrite formula (\ref{b_k}) as

~
\[
b_{k}=b_{k}^{(1)}+b_{k}^{(2)}-b_{k}^{(3)},
\]
where
\[
b_{k}^{(1)}=\dfrac{\lambda^{-2}}{P\left(\lambda_{k}\right)},
\]

\[
b_{k}^{(2)}=\lambda^{-1}\dfrac{1-p^{-\beta}}{1-p^{-\left(\beta-1\right)r}}\dfrac{R\left(\lambda_{k},\alpha,\beta,r\right)}{P\left(\lambda_{k}\right)},
\]

\[
b_{k}^{(3)}=\lambda^{-1}\dfrac{\left(1-p^{-1}\right)p^{-\left(\beta-1\right)r}}{1-p^{-\left(\beta-1\right)r}}\dfrac{R\left(\lambda_{k},\alpha,1,r\right)}{P\left(\lambda_{k}\right)},
\]

\[
P\left(\lambda_{k},\alpha\right)=\dfrac{dJ\left(s\right)}{ds}|_{s=-\lambda_{k}}
\]

\[
=\left(1-p^{-1}\right)\sum_{n=0}^{\infty}\dfrac{p^{-n}}{\left(p^{-\alpha n}-\lambda_{k}\right)^{2}},
\]

\[
R\left(\lambda_{k},\alpha,\beta,r\right)={\displaystyle \sum_{n=0}^{r-1}}\dfrac{p^{-\beta n}}{-\lambda_{k}+p^{-\alpha n}}.
\]
Using the inequalities \begin{widetext}
\[
{\displaystyle \sum_{n=0}^{\infty}\dfrac{p^{-n}}{\left(p^{-\alpha n}-\lambda_{k}\right)^{2}}>\dfrac{p^{-k-1}}{\left(p^{-\alpha\left(k+1\right)}-\lambda_{k}\right)^{2}}=\dfrac{1}{p}\dfrac{p^{\left(2\alpha-1\right)k}}{\left(1-p^{-\alpha}\right)^{2}\delta_{k}^{2}}},
\]

\[
{\displaystyle \sum_{n=0}^{\infty}\dfrac{p^{-n}}{\left(p^{-\alpha n}-\lambda_{k}\right)^{2}}<\sum_{n=0}^{k}\dfrac{p^{-n}}{\left(p^{-\alpha n}-\lambda_{n-1}\right)^{2}}}+\dfrac{p^{-k-1}}{\left(p^{-\alpha\left(k+1\right)}-\lambda_{k}\right)^{2}}+{\displaystyle \sum_{n=k+2}^{\infty}\dfrac{p^{-n}}{\left(p^{-\alpha\left(k+1\right)}-\lambda_{k}\right)^{2}}}
\]

\[
<\dfrac{p^{\left(2\alpha-1\right)k}}{\left(1-p^{-\alpha}\right)^{2}\delta_{k}^{2}}\dfrac{1}{p\left(p-1\right)}\dfrac{p^{2\alpha}-1}{p^{2\alpha-1}-1}
\]
\end{widetext}

and, taking into account (\ref{App_A_Result}), we can obtain the
following structure of the upper and lower bounds for $P\left(\lambda_{k},\alpha\right)$:

\begin{equation}
U_{1}p^{\left(2\alpha-1\right)k}<P\left(\lambda_{k},\alpha\right)<{\displaystyle U_{2}p^{\left(2\alpha-1\right)k}}\label{P}
\end{equation}
where the coefficients $U_{i}$ are independent of $k$.

To find the structure of the estimates $R\left(\lambda_{k},\alpha,\beta,r\right)$
for $k\leqslant r-1$, it is convenient to single out the term with
$n=k$ in the sum:
\[
R\left(\lambda_{k},\alpha,\beta,r\right)={\displaystyle \sum_{n=0}^{k-1}\dfrac{p^{-\beta n}}{p^{-\alpha n}-\lambda_{k}}}
\]
\begin{equation}
+\dfrac{p^{-\beta k}}{p^{-\alpha k}-\lambda_{k}}-{\displaystyle \sum_{n=k+1}^{r-1}}\dfrac{p^{-\beta n}}{\lambda_{k}-p^{-\alpha n}}.\label{R_lambda}
\end{equation}
It follows from (\ref{R_lambda}) that

\[
R\left(\lambda_{k},\alpha,\beta,r\right)<{\displaystyle \sum_{n=0}^{k-1}\dfrac{p^{-\beta n}}{p^{-\alpha n}-\lambda_{k}}}+\dfrac{p^{-\beta k}}{p^{-\alpha k}-\lambda_{k}}
\]

\[
<{\displaystyle \sum_{n=0}^{k-1}\dfrac{p^{\left(\alpha-\beta\right)n}}{\left(1-p^{-\alpha}\right)\left(1-\delta_{k}\right)}}+\dfrac{p^{\left(\alpha-\beta\right)k}}{\left(1-p^{-\alpha}\right)\left(1-\delta_{k}\right)}
\]

\[
<\dfrac{1}{\left(1-p^{-\alpha}\right)\left(1-\delta_{k}\right)}\left({\displaystyle \dfrac{1-p^{\left(\alpha-\beta\right)\left(k+1\right)}}{1-p^{\left(\alpha-\beta\right)}}}\right).
\]

\[
R\left(\lambda_{k},\alpha,\beta,r\right)>\sum_{n=0}^{k-1}p^{\left(\alpha-\beta\right)n}
\]

\[
+\dfrac{p^{-\beta k}}{p^{-\alpha k}-\lambda_{k}}-{\displaystyle \sum_{n=k+1}^{r-1}}\dfrac{p^{-\beta n}}{\lambda_{k}-p^{-\alpha n}}
\]

\[
>\dfrac{p^{\left(\alpha-\beta\right)k}-1}{p^{\alpha-\beta}-1}+\dfrac{p^{\left(\alpha-\beta\right)k}}{\left(1-p^{-\alpha}\right)\left(1-\delta_{k}\right)}
\]
\[
-\dfrac{p^{\left(\alpha-\beta\right)k}}{\left(1-p^{-\alpha}\right)\delta_{k}\left(p^{\beta}-1\right)}+\dfrac{p^{-\beta r}p^{\alpha k}}{\left(1-p^{-\alpha}\right)\delta_{k}\left(p^{\beta}-1\right)}.
\]
This implies the following structure of estimates:

\begin{equation}
V_{1}p^{\left(\alpha-\beta\right)k}<R\left(\lambda_{k},\alpha,\beta,r\right)<V_{2}p^{\left(\alpha-\beta\right)k}\label{R1}
\end{equation}
for $k\leqslant r-1$ and $\alpha>\beta$ and

\begin{equation}
V_{3}<R\left(\lambda_{k},\alpha,\beta,r\right)<V_{4}\label{R2}
\end{equation}
and for $k\leqslant r-1$ and $\alpha<\beta$, the coefficients $V_{i}$
in (\ref{R1}) and (\ref{R2}) being independent of $k$ and $r$.
Similarly, we can show that, for $k>r-1$,

\begin{equation}
\left(p^{\left(\alpha-\beta\right)r}-1\right)W_{1}<R\left(\lambda_{k},\alpha,\beta,r\right)<\left(1-p^{-\beta r}\right)p^{\alpha r}W_{2};\label{R3}
\end{equation}
in (\ref{R3}), $W_{i}$ are independent of $k$ and $r$.

Using (\ref{P}), (\ref{R1}), (\ref{R2}), and (\ref{R3}), we find
the structure of upper and lower bounds for $b_{k}^{(1)}$, $b_{k}^{(2)}$
and $b_{k}^{(3)}$:
\begin{equation}
g^{(1)}p^{-\left(2\alpha-1\right)k}<b_{k}^{(1)}<h^{(1)}p^{-\left(2\alpha-1\right)k}\label{b_1}
\end{equation}
for any $k$ and $\alpha$,
\begin{equation}
g^{(2)}p^{-\left(\alpha+\beta-1\right)k}<b_{k}^{(2)}<h^{(2)}p^{-\left(\alpha+\beta-1\right)k},\label{b_2_(1)}
\end{equation}
\begin{equation}
g^{(3)}p^{-\left(\beta-1\right)r}p^{-\left(\alpha+\beta-1\right)k}<b_{k}^{(3)}<h^{(3)}p^{-\left(\beta-1\right)r}p^{-\left(\alpha+\beta-1\right)k},\label{b_3_(1)}
\end{equation}
for $k\leqslant r-1$ and $\alpha>\beta$,
\begin{equation}
g^{(2)}p^{-\left(2\alpha-1\right)k}<b_{k}^{(2)}<h^{(2)}p^{-\left(2\alpha-1\right)k},\label{b_2_(2)}
\end{equation}
\begin{equation}
g^{(3)}p^{-\left(\beta-1\right)r}p^{-\left(2\alpha-1\right)k}<b_{k}^{(3)}<h^{(1)}p^{-\left(\beta-1\right)r}p^{-\left(2\alpha-1\right)k},\label{b_3_(2)}
\end{equation}
for $k\leqslant r-1$ and $\alpha<\beta$, and
\[
g^{(2)}\left(p^{\left(\alpha-\beta\right)r}-1\right)p^{-\left(2\alpha-1\right)k}<b_{k}^{(2)}
\]
\begin{equation}
<h^{(2)}\left(1-p^{-\beta r}\right)p^{\alpha r}p^{-\left(2\alpha-1\right)k}\label{b_2_(3)}
\end{equation}
\[
g^{(3)}\left(p^{\left(\alpha-\beta\right)r}-1\right)p^{-\left(2\alpha-1\right)k}<b_{k}^{(3)}
\]
\begin{equation}
<h^{(1)}\left(1-p^{-\beta r}\right)p^{\alpha r}p^{-\left(2\alpha-1\right)k},\label{b_3_(3)}
\end{equation}
for $k>r-1$. The coefficients $g^{(i)}$, $h^{(i)}$, $i=1,2,3$,
are independent of $r$ and $k$, and their explicit form is unimportant
for us.

\section{Asymptotic estimate of the series $S\left(t\right)={\displaystyle \sum_{n=0}^{\infty}a^{-n}\exp\left(-b^{-n}t\right)}$
and $S\left(t,r\right)={\displaystyle \sum_{n=0}^{r}a^{-n}\exp\left[-b^{-n}t\right]}$}

\label{A_C}

\textbf{Theorem 1.} \textit{Suppose given a series
\[
S\left(t\right)={\displaystyle \sum_{n=0}^{\infty}a^{-n}\exp\left(-b^{-n}t\right)}
\]
with $a>1$ and $b>1$. Then, for any $0<M<1,$ there exists a $T>0$
such that the following inequalities hold for $t>T:$}

\textit{
\[
\dfrac{a^{-1}}{\ln b}t^{-z}{\displaystyle \left(\Gamma\left(z\right)-\left(bt\right)^{z-1}\exp\left(-bt\right)\right)<}
\]
}

\textit{\emph{
\begin{equation}
S\left(t\right)<\dfrac{a}{\ln b}t^{-z}\left(\Gamma\left(z\right)-Mt^{z-1}\exp\left(-t\right)\right),\label{AS_T_1}
\end{equation}
}}\emph{ where $z=\dfrac{\log a}{\log b}$. }

\textbf{Proof.} Since $a^{-x}$ is a monotonically decreasing function
and $\exp\left(-b^{-x}t\right)$ is a monotonically increasing function,
it follows that the following inequality holds for $x\in\left[n,\:n+1\right]$:
\[
a^{-x}\exp\left[-b^{-\left(x-1\right)}t\right]\leqslant a^{-n}\exp\left[-b^{-n}t\right]
\]
\begin{equation}
\leqslant a^{-\left(x-1\right)}\exp\left[-b^{-x}t\right].\label{AS_1}
\end{equation}
Integrating this inequality over the interval $\left[n,\:n+1\right]$
with respect to $x$ and then summing over $n$ from $0$ to $\infty$,
we obtain
\[
{\displaystyle a^{-1}\intop_{0}^{\infty}}a^{-\left(x-1\right)}\exp\left[-b^{-\left(x-1\right)}t\right]dx\leqslant S\left(t\right)
\]
\begin{equation}
\leqslant{\displaystyle a\intop_{0}^{\infty}}a^{-x}\exp\left[-b^{-x}t\right]dx.\label{AS_2}
\end{equation}
In the left and right integrals in (\ref{AS_2}), we make the changes
$x\rightarrow y=b^{-\left(x-1\right)}t$ and $x\rightarrow y=b^{-x}t$,
respectively. Then we obtain

\begin{equation}
\dfrac{a^{-1}}{\ln b}t^{-z}{\displaystyle \gamma\left(z,bt\right)\leq S\left(t\right)\leq\dfrac{a}{\ln b}t^{-z}\gamma\left(z,t\right)},\label{<S<}
\end{equation}
where $\gamma\left(z,x\right)=\intop_{0}^{x}y^{z-1}\exp\left(-y\right)dy$
is the incomplete gamma function.

Using the asymptotic expansion of $\gamma\left(z,t\right)$ as $t\rightarrow\infty$,
\[
\gamma\left(z,t\right)=\Gamma\left(z\right)-t^{z-1}\exp\left(-t\right)
\]
\begin{equation}
-t^{z-1}\exp\left(-t\right)\left\{ {\displaystyle \sum_{n=1}^{m-1}\dfrac{\left(-1\right)^{n}}{t^{n}}\dfrac{\Gamma\left(1-z+n\right)}{\Gamma\left(1-z\right)}+O\left(t^{-m}\right)}\right\} ,\label{AS_gamma}
\end{equation}
we can write \begin{widetext}
\[
\gamma\left(z,t\right)=\Gamma\left(z\right)-Mt^{z-1}\exp\left(-t\right)-t^{z-1}\exp\left(-t\right)\left\{ \left(1-M\right)+{\displaystyle \sum_{n=1}^{m-1}\dfrac{\left(-1\right)^{n}}{t^{n}}\dfrac{\Gamma\left(1-z+n\right)}{\Gamma\left(1-z\right)}+O\left(t^{-m}\right)}\right\} ,
\]
\end{widetext}

where $0<M<1$ is an arbitrary number. For sufficiently large $T$,
the expression in the curly brackets $\left\{ \cdots\right\} $ is
positive for $t>T$, and we can write
\begin{equation}
\gamma\left(z,t\right)<\Gamma\left(z\right)-Mt^{z-1}\exp\left(-t\right).\label{gamma<}
\end{equation}
Completely analogously we can obtain
\begin{equation}
\gamma\left(z,bt\right)>\Gamma\left(z\right)-t^{z-1}b^{z-1}\exp\left(-bt\right).\label{gamma>}
\end{equation}
Inequalities (\ref{gamma<}), (\ref{gamma>}), and (\ref{<S<}) imply
the assertion of Theorem 1.

\textbf{Theorem 2.} \textit{Suppose given a series
\[
S\left(t,r\right)={\displaystyle \sum_{n=0}^{r}a^{-n}\exp\left(-b^{-n}t\right)}
\]
with $a>1$ and $b>1$. Then, for any $0<M<1$ and $0<N<1$, there
exist $T>0$ and $\delta>0$ such that the following inequalities
hold for any $t$ and $r$ satisfying $t>T$ and $b^{-r}t<\delta:$}
\begin{widetext} \textit{
\begin{equation}
\dfrac{a^{-1}}{\ln b}t^{-z}{\displaystyle \left(\Gamma\left(z\right)-\left(bt\right)^{z-1}\exp\left(-bt\right)-\left(b^{-r}t\right)^{z}\dfrac{1}{z}\right)<}S\left(t,r\right)<\dfrac{a}{\ln b}t^{-z}\left(\Gamma\left(z\right)-Mt^{z-1}\exp\left(-t\right)-N\left(b^{-r}t\right)^{z}\dfrac{b}{z}^{-z}\right).\label{AS_T_2}
\end{equation}
} \end{widetext}

\textbf{Proof. } Integrating inequality (\ref{AS_1}) over the interval
$\left[n,\:n+1\right]$ with respect to $x$ and then summing over
$n$ from $0$ to $r$, we obtain \begin{widetext}
\begin{equation}
{\displaystyle a^{-1}\intop_{0}^{r}}a^{-\left(x-1\right)}\exp\left[-b^{-\left(x-1\right)}t\right]dx\leqslant S\left(t,r\right)\leqslant{\displaystyle a\intop_{0}^{r}}a^{-x}\exp\left[-b^{-x}t\right]dx.\label{AS_3}
\end{equation}
\end{widetext}

Making the changes $x\rightarrow y=b^{-\left(x-1\right)}t$ and $x\rightarrow y=b^{-x}t$,
respectively, in the left and right integrals in (\ref{AS_3}) and
denoting $z=\dfrac{\log a}{\log b}$, we obtain
\begin{equation}
\dfrac{a^{-1}}{\ln b}t^{-z}{\displaystyle \gamma\left(z,b^{-r}t,bt\right)\leq S\left(t,r\right)\leq\dfrac{a}{\ln b}t^{-z}\gamma\left(z,b^{-r-1}t,t\right),}\label{AS_4}
\end{equation}
where $\gamma\left(z,x_{1}x_{2}\right)=\intop_{x_{1}}^{x_{2}}y^{z-1}\exp\left(-y\right)dy=\gamma\left(z,x_{2}\right)-\gamma\left(z,x_{1}\right)$.

Consider
\begin{equation}
\gamma\left(z,b^{-r-1}t,t\right)=\gamma\left(z,t\right)-\gamma\left(z,b^{-r-1}t\right).\label{gamma_z}
\end{equation}

For sufficiently large $t$, i.e., for $t>T$, for the first term
of (\ref{gamma_z}) we have (\ref{gamma<}). For the second term of
(\ref{gamma_z}), we have \begin{widetext}
\[
\gamma\left(z,b^{-r-1}t\right)={\displaystyle \intop_{0}^{tb^{-r-1}}y^{z-1}\exp\left(-y\right)dy}={\displaystyle \intop_{0}^{tb^{-r-1}}y^{z-1}{\displaystyle \sum_{n=0}^{\infty}\dfrac{\left(-1\right)^{n}}{n!}y^{n}dy}}
\]

\[
\dfrac{b^{-z}}{z}\left(\dfrac{t}{b^{r}}\right)^{z}-{\displaystyle \sum_{n=1}^{\infty}}\dfrac{\left(-1\right)^{n+1}}{n!}\dfrac{b^{-z-n}}{z+n}\left(\dfrac{t}{b^{r}}\right)^{z+n}=N\dfrac{b^{-z}}{z}\left(\dfrac{t}{b^{r}}\right)^{z}+\left\{ \left(1-N\right)\dfrac{b^{-z}}{z}\left(\dfrac{t}{b^{r}}\right)^{z}-{\displaystyle \sum_{n=1}^{\infty}}\dfrac{\left(-1\right)^{n+1}}{n!}\dfrac{b^{-z-n}}{z+n}\left(\dfrac{t}{b^{r}}\right)^{z+n}\right\} ,
\]
\end{widetext}

where $0<N<1$ is an arbitrary number. For sufficiently small $\delta$
for $tb^{-r}<\delta$, the expression in the curly brackets $\left\{ \cdots\right\} $
is positive, and we can write

\begin{equation}
\gamma\left(z,b^{-r-1}t\right)>N\dfrac{b^{-z}}{z}\left(\dfrac{t}{b^{r}}\right)^{z}.\label{gamma_r>}
\end{equation}

It follows from (\ref{gamma<}) and (\ref{gamma_r>}) that
\begin{equation}
\gamma\left(z,b^{-r-1}t,t\right)<\Gamma\left(z\right)-Mt^{z-1}\exp\left(-t\right)-N\dfrac{b^{-z}}{z}\left(\dfrac{t}{b^{r}}\right)^{z}.\label{gamma_r_up}
\end{equation}

Similarly we can prove that
\begin{equation}
\gamma\left(z,b^{-r}t,bt\right)>\Gamma\left(z\right)-\left(bt\right)^{z-1}\exp\left(-bt\right)-\left(b^{-r}t\right)^{z}\dfrac{1}{z}.\label{gamma_r_down}
\end{equation}

Inequalities (\ref{gamma_r_up}), (\ref{gamma_r_down}), and (\ref{AS_4})
imply the assertion of Theorem 2.



\end{document}